\documentclass{svmult}

\usepackage{makeidx}         
\usepackage{graphicx}        
\usepackage{multicol}        
\usepackage[bottom]{footmisc}
\usepackage{natbib}
\usepackage{url}
\usepackage{xcolor}
\usepackage{subfig}
\usepackage{siunitx}
\usepackage[utf8]{inputenc}

\renewcommand{\PackageWarningNoLine}[2]{}

\newcommand{\coloneq}{\mathrel{\mathop:}=}

\usepackage{amsmath}
\usepackage{amsfonts}
\usepackage{array}

\begin{document}

\title*{Parareal convergence for 2D unsteady flow around a cylinder}

\author{Andreas Kreienbuehl\inst{1}\and
Arne Naegel\inst{2}\and
Daniel Ruprecht\inst{3} \and
Andreas Vogel\inst{2} \and
Gabriel Wittum\inst{2} \and
Rolf Krause\inst{1}
}
\authorrunning{Kreienbuehl et al.}

\institute{
$^1$Institute of Computational Science, Faculty of Informatics, Universit{\`a} della Svizzera italiana, Lugano, Switzerland
\and
$^2$Goethe-Center for Scientific Computing, Goethe-University Frankfurt am Main, Kettenhofweg 139, DE-60325 Frankfurt am Main, Germany
\and
$^3$School of Mechanical Engineering, University of Leeds, Woodhouse Lane, Leeds LS2 9JT, UK.
}
%
%
\maketitle

\abstract{
In this technical report we study the convergence of Parareal for 2D incompressible flow around a cylinder for different viscosities.
Two methods are used as fine integrator: backward Euler and a fractional step method.
It is found that Parareal converges better for the implicit Euler, likely because it under-resolves the fine-scale dynamics as a result of numerical diffusion.
}


\section{Introduction}\label{s:introduction}

The potential of parallel-in-time integration methods to increase the degree of concurrency in the numerical solution of time-dependent partial differential equations has been widely acknowledged, \textit{e.g.} in the report \emph{Applied Mathematics Research for Exascale Computing} by \cite{DongarraEtAl2014}.
A variety of different methods exists, see \textit{e.g.} the review by~\cite{Gander2015_Review}, and principle efficiency of parallel-in-time integration in large- and extreme-scale parallel computations has been demonstrated, \textit{e.g.} in ~\cite{SpeckEtAl2012,RuprechtEtAl2013_SC}, and~\cite{Neumueller2014}.

Many problems in computational fluid dynamics require massive computational capacities and suffer from long solution times.
Exploring the potential of parallel-in-time methods to speed up such simulations can therefore be a beneficial endeavour.
The performance of the parallel-in-time integration method Parareal, introduced by~\cite{LionsEtAl2001}, when applied to the Navier-Stokes equations has been a topic of research since shortly after its introduction.
First studies have been conducted by~\cite{Trindade2004,Trindade2006}, including reports of speedup for an \texttt{MPI} implementation; laminar flow around a cylinder is used as benchmark problem and it is shown that Parareal can correctly reproduce the Nusselt number.
\cite{FischerEtAl2005} investigate Parareal for spatial discretisations based on finite and spectral element methods, and discuss using fewer spatial degrees-of-freedom for the coarse integrator.
The performance of Parareal for simulations of non-Newtonian fluids has been investigated by~\cite{Celledoni2009}.
Finally, parallel scaling of Parareal for 3D unsteady flow is investigated by~\cite{CroceEtAl2014} on up to \num{2048} cores.

Based on predictions from linear stability analysis by~\cite{GanderVandewalle2007}, it has been shown by~\cite{SteinerEtAl2015}) that convergence of Parareal deteriorates as the Reynolds number increases.
However, the studies only analysed a rather simple driven cavity problem, which eventually approaches a steady-state and thus may underestimate the problem because of weak transient dynamics towards the end of the simulation.
In this report, we continue this investigation for a different, more complex benchmark involving unsteady flow around a cylinder.
It was introduced by~\cite{SchaeferAndTurek1996a} as the case $2$D-$3$ and further analyzed by \textit{e.g.} \cite{John2004a}.
Eventually, the here presented benchmarks will be extended to a comprehensive exploration of Parareal's performance for 3D flow, including a study of the influence of spatial resolution.

\section{Parareal and model problem}\label{s:parareal}

\subsection{Parareal}

Parareal parallelises the solution of initial value problems
\begin{equation}
	u_t = f(t,u(t)),\quad u(0) = u_0, \quad t \in [0,T],
\end{equation}
by decomposing the time domain $[0,T]$ into time slices $[t_{j-1}, t_{j}]$, $j=1,\ldots,N_{\text{pr}}$ with $N_{\text{pr}}$ equal to the number of processing units.
It then iterates between two time integration methods: a coarse integrator $\mathcal{C}$ used to serially propagate corrections, which has to be computationally cheap, and an accurate integrator $\mathcal{F}$ run in parallel.
The Parareal iteration reads
\begin{equation}
	u^{k+1}_{j+1} = \mathcal{C}(u^{k+1}_j) + \mathcal{F}(u^k_j) - \mathcal{C}(u^k_j),
\quad j=1,\ldots,N_{\text{pr}},
\end{equation}
with $k$ being the iteration index and $u_j \approx u(t_j)$.
Note how the computationally expensive computation of the fine method can be done concurrently for all time slices.
A detailed presentation including a theoretical model for projected speedup is given \textit{e.g.} by~\cite{Minion2010}.

\subsection{Model problem}
As model problem, we consider the Navier-Stokes equations
\begin{equation}
(\partial_t+\boldsymbol{U}\boldsymbol{\nabla})\boldsymbol{U}=-\frac{\boldsymbol{\nabla}p}{\varrho}+\nu\boldsymbol{\nabla}^2\boldsymbol{U}
\end{equation}
for an incompressible fluid, \textit{i.e.}\ for a fluid with $\boldsymbol{\nabla}\boldsymbol{U}=0$, at density $\varrho\coloneq1\ (\text{kg}/\text{m}^2)$. We focus on the benchmark problem defined in~\cite{SchaeferAndTurek1996a} as $2$D-$3$, which is for \textit{unsteady} flow around a cylinder in two dimensions, \textit{i.e.}~in $2$D (see also~\cite{John2004a}). 
We make use of the definitions
\begin{equation}
	\boldsymbol{\nabla}\coloneq(\partial_x,\partial_y)^{\text{T}},\qquad\boldsymbol{U}\coloneq\boldsymbol{U}(t,x,y),\qquad\boldsymbol{U}\coloneq(u,v)^{\text{T}}.
\end{equation}
The Reynolds number for a cylinder with diameter $d$ (the reference length) located inside a square cuboid with longest edge along the $x$-coordinate is
\begin{equation}
	R_d\coloneq\bar{u}_{\text{in}}\frac{d}{\nu},
\end{equation}
where the reference velocity $\bar{u}_{\text{in}}$ is chosen to be the \textit{mean} velocity of inflow in $x$-direction and $\nu$ the kinematic viscosity. Notice that $R_d$ can be time dependent, as is the case for the problem considered here.

In the $2$D case the \textit{mean} velocity is
\begin{equation}\label{e:mean_2d_velocity}
	\bar{u}(t)\coloneq\int_0^h\frac{u(t,0,y)}{h}\ \text{\text{d}}y=\frac{2}{3}u\left(t,0,\frac{h}{2}\right),\qquad\bar{v}(t)\coloneq0,
\end{equation}
where $h\coloneq0.41\ (\text{m})$. In the $2$D-$3$ case we have the \textit{inflow} velocity
\begin{equation}
	u_{\text{in}}(t,0,y)\coloneq4u^{\text{x}}_{\text{in}}\sin\left(\frac{\pi}{8}t\right)\frac{y(h-y)}{h^2},\qquad v_{\text{in}}(t,0,y)\coloneq0,
\end{equation}
for which Equation~\eqref{e:mean_2d_velocity} is valid.
We choose $u^{\text{x}}_{\text{in}}\coloneq3/2\ (\text{m}/\text{s})$ so that
\begin{equation}
	R_d(t)=\sin\left(\frac{\pi}{8}t\right)\frac{d}{\nu}\in\left[0,\frac{d}{\nu}\right].
\end{equation}
Thus, setting $d\coloneq0.1\ (\text{m})$, it follows for $\nu\in\{0.1,0.01,0.001\}\ (\text{m}/\text{s})$ that
\begin{equation}
	R^{\text{x}}_d\in\{1,10,100\}
\end{equation}
defines the maximum-over-time Reynolds number for the chosen $\nu$.

In this report, the primary goal is to outline the performance of Parareal for the $2$D-$3$ benchmark problem for the three mentioned viscosities, \textit{i.e.}\ ranges of Reynolds numbers.

\subsection{Implementation details}

The governing equations were implemented using Q2-Q1 finite elements in the \texttt{UG4} software toolbox (see \cite{VogelUG42013, VogelDiss2014}). For the parallelisation in time via Parareal, we used the library \texttt{Lib4PrM}, which was first applied in~\cite{KreienbuehlEtAl2015a}.\footnote{It can be obtained by cloning the \texttt{Git} repository \url{https://scm.ti-edu.ch/repogit/lib4prm}.}

\section{Results}\label{s:results}

Instead of measuring convergence of Parareal by comparing the discretisation error with the defect as discussed \textit{e.g.} by~\cite{ArteagaEtAl2015}, we focus here on how well Parareal reproduces important characteristic numbers of the dynamics, namely the \emph{drag coefficient} $C_{\text{dr}}$, the \emph{lift coefficient} $C_{\text{li}}$ and the \emph{pressure difference} $\Delta_p$ between the \textit{front} and \textit{end} point of the cylinder over time.
These parameters are respectively defined as follows:
\begin{equation}
	C_{\text{dr}}\coloneq\frac{2F_{\text{dr}}}{\varrho\bar{u}^2_{\text{in}}d},\qquad C_{\text{li}}\coloneq\frac{2F_{\text{li}}}{\varrho\bar{u}^2_{\text{in}}d},\qquad\Delta_p\coloneq p_{\text{fr}}-p_{\text{en}},
\end{equation}
where $F_{\text{dr}}$ is the drag force and $F_{\text{li}}$ the lift force, and $p_{\text{fr}}$ together with $p_{\text{en}}$ define the pressure at the front and end of the cylinder. 
Again, we assume here that the density is $1\ (\text{kg}/\text{m}^2)$ and set $t\in[0,8]\ (\text{s})$ as time domain.

For $\nu=0.001\ (\text{m}/\text{s})$, \cite{SchaeferAndTurek1996a} report on a \textit{maximum}-over-time drag coefficient of $C^{\text{ma}}_{\text{dr}}\approx2.9500\pm0.0200$, a \textit{maximum}-over-time lift coefficient of $C^{\text{ma}}_{\text{li}}\approx0.4800\pm0.0100$, and a \textit{final} pressure difference at $t=8\ (\text{s})$ of $\Delta^{\text{fi}}_p\approx-0.1100\pm0.0050\ (\text{kg}/\text{s}^2)$.

\subsection{Numerical setup}\label{s:numerics}

We use $N_{\text{pr}}\in\{2,4,8,16\}$ processors without parallelization in space and with $13,\!212$ spatial degrees-of-freedom for both the fine and coarse level. 
For each $\nu$, we consider the following two Parareal solvers ``S'' comprised of a coarse $\text{M}_{\text{co}}$ and fine $\text{M}_{\text{fi}}$ serial time integration method as well as number of time steps $N_{\text{co}}$ and $N_{\text{fi}}$:
\begin{align*}
	(\text{M}_{\text{co}},N_{\text{co}})\times(\text{M}_{\text{fi}},N_{\text{fi}})&=(\text{IE},16)\times(\text{IE},32),\tag{$\text{S}_1$}\\
		(\text{M}_{\text{co}},N_{\text{co}})\times(\text{M}_{\text{fi}},N_{\text{fi}})&=(\text{IE},16)\times(\text{FS},32),\tag{$\text{S}_2$}
\end{align*}
where ``IE'' stands for implicit Euler (first-order) and ``FS'' for fractional step (second-order).
Errors in the three physical quantities discussed above are measured by
\begin{equation}\label{e:error_norm}
	E_{\text{ph}}\coloneq\frac{\|u^{\text{pa}}_{\text{ph}}-u^{\text{fi}}_{\text{ph}}\|_{\text{ti}}}{\|u^{\text{fi}}_{\text{ph}}\|_{\text{ti}}},
\end{equation}
where the parameter ``ph'' is in $\{$dr,li,$p\}$ for \textit{drag} and \textit{lift} coefficient or \textit{pressure} difference. 
We use the $l_2$-norm over the solutions at the end of all time-slices
\begin{equation}
\|u\|^2_{\text{ti}}\coloneq\frac{8}{N_{\text{pr}}}\sum_{n=1}^{N_{\text{pr}}}|u(t^n)|^2
\end{equation}
weighted by the time slice length $8/N_{\text{pr}}$.
	
\subsection{Problem dynamics}
\begin{figure}[!t]
	\centering
	\includegraphics[width=\textwidth]{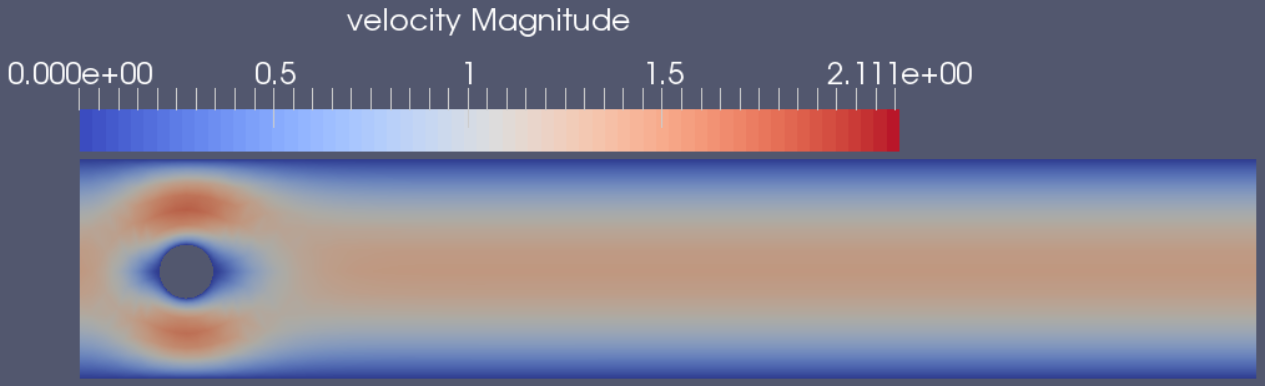}\\\vspace{0.25\baselineskip}
	\includegraphics[width=\textwidth]{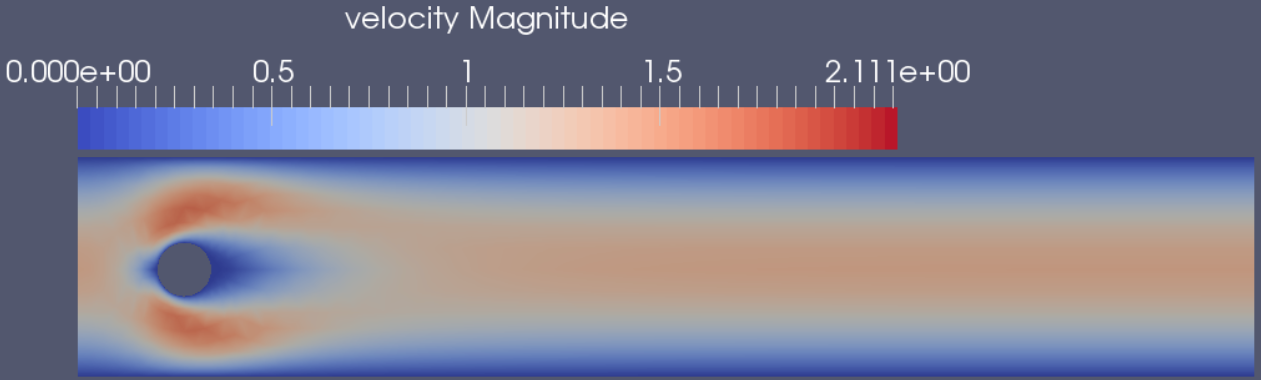}\\\vspace{0.25\baselineskip}
	\includegraphics[width=\textwidth]{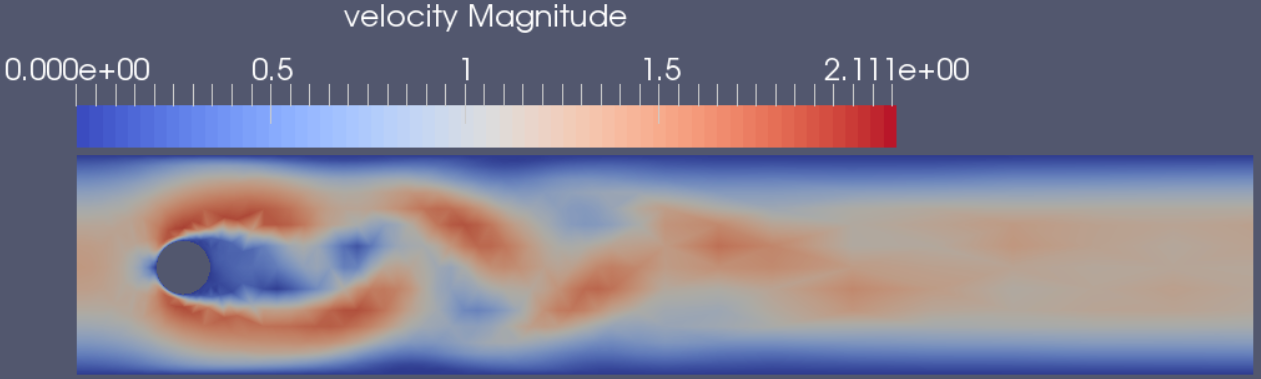}
	\caption{Flow field at $t=5.25\ (\text{s})$ for three different viscosities: $\nu=0.1\ (\text{m}/\text{s})$ at the top, $\nu=0.01\ (\text{m}/\text{s})$ in the middle, and $\nu=0.001\ (\text{m}/\text{s})$ at the bottom.}
	\label{f:flowfield}
\end{figure}

\begin{figure}[!tph]
\centering	

\includegraphics{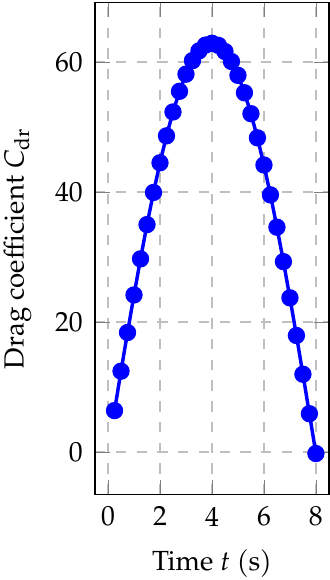}\hfill
\includegraphics{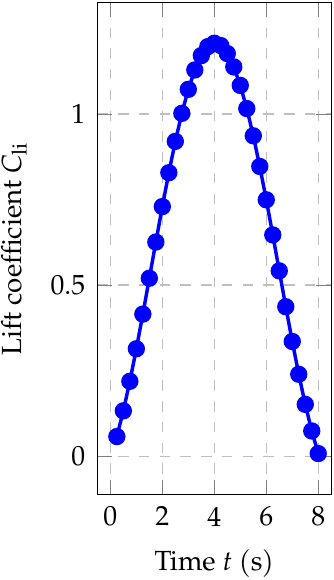}\hfill
\includegraphics{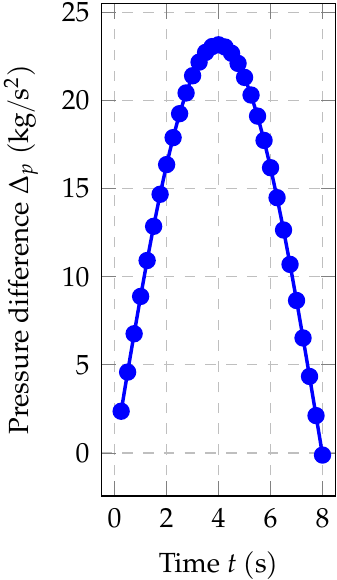}

\includegraphics{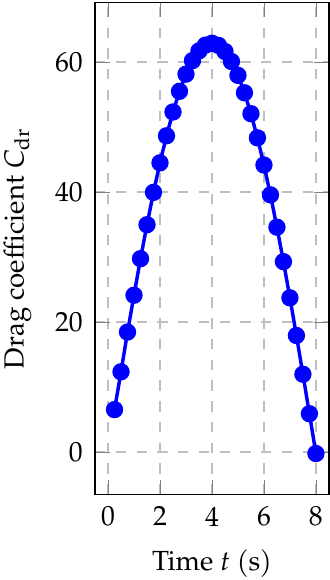}\hfill
\includegraphics{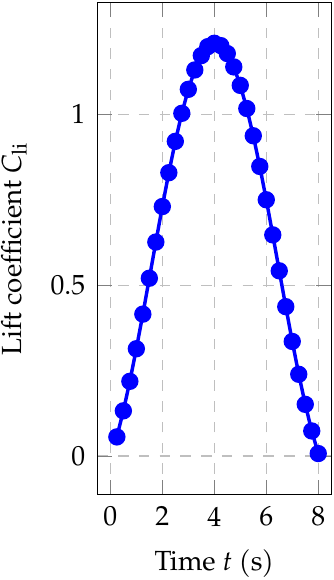}\hfill
\includegraphics{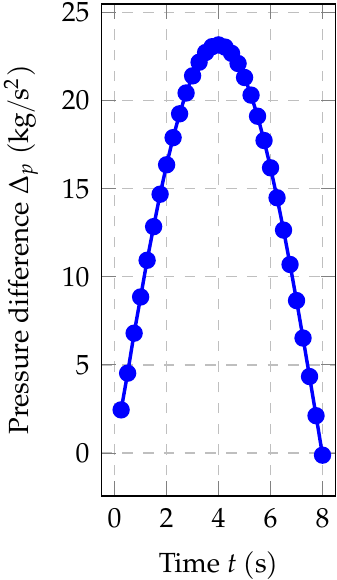}

\caption{Drag and lift coefficient, and pressure difference for $\nu=0.1\ (\text{m}/\text{s})$ for $S_1$ (top) and $S_2$ (bottom). The curves match the values from the reference simulation (not shown).}
\label{f:charnumbers-nu0.1}
\end{figure}

\begin{figure}[!tp]
\centering
\includegraphics{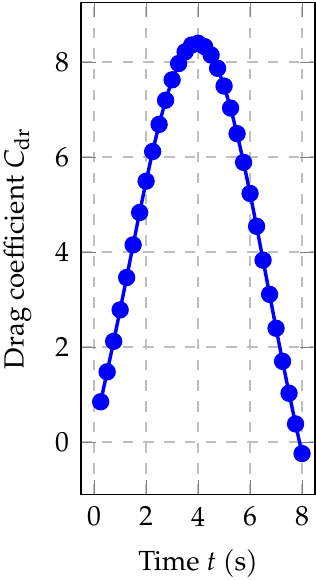}\hfill
\includegraphics{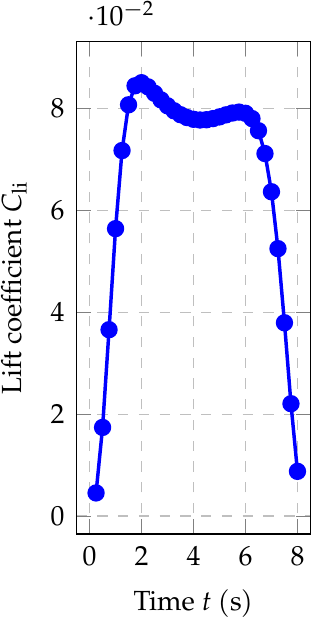}\hfill
\includegraphics{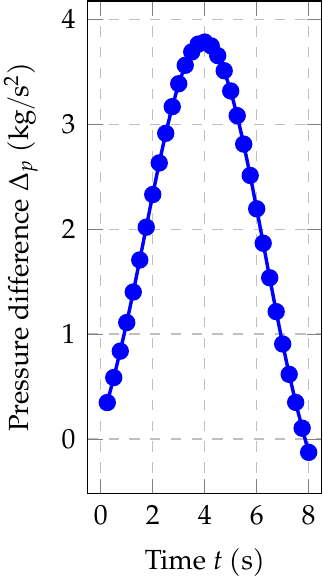}
\includegraphics{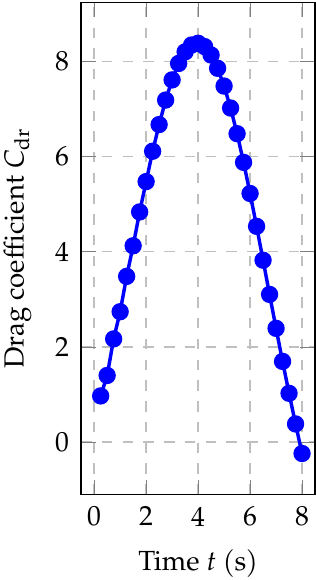}\hfill
\includegraphics{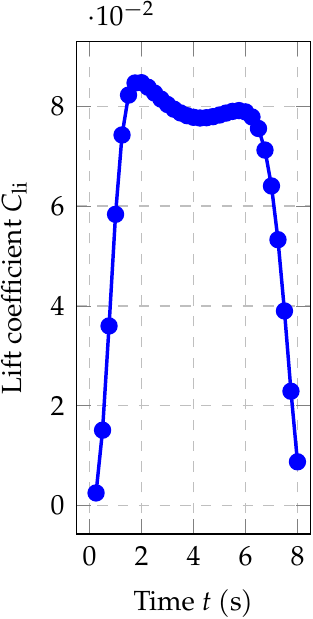}\hfill
\includegraphics{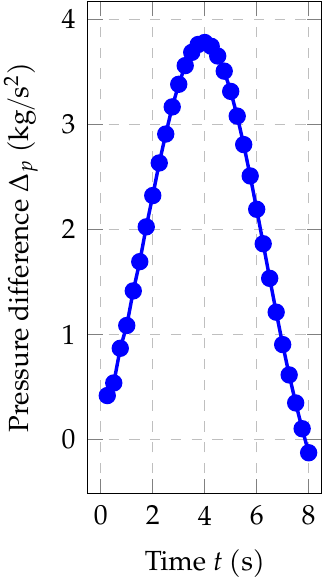}
\caption{Drag and lift coefficient, and pressure difference for $\nu=0.01\ (\text{m}/\text{s})$ for reference (top), $S_1$ (middle) and $S_2$ (bottom).}
\label{f:charnumbers-nu0.01}
\end{figure}

\begin{figure}[!tph]
\centering
\includegraphics{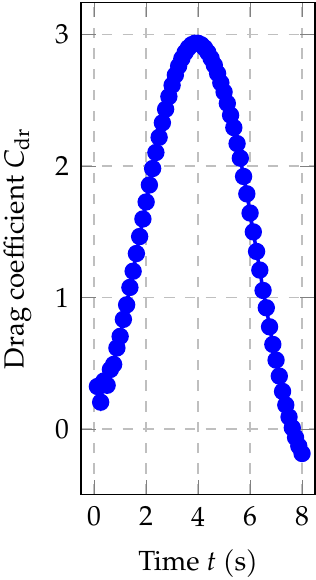}\hfill
\includegraphics{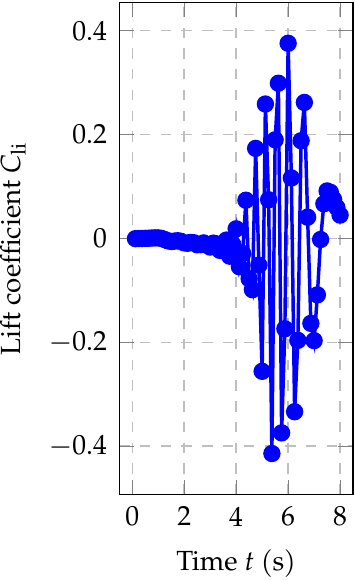}\hfill
\includegraphics{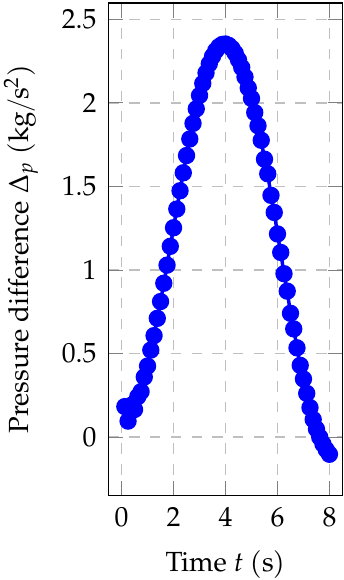}
\includegraphics{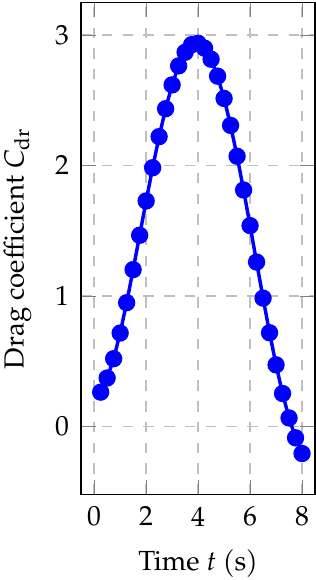}\hfill
\includegraphics{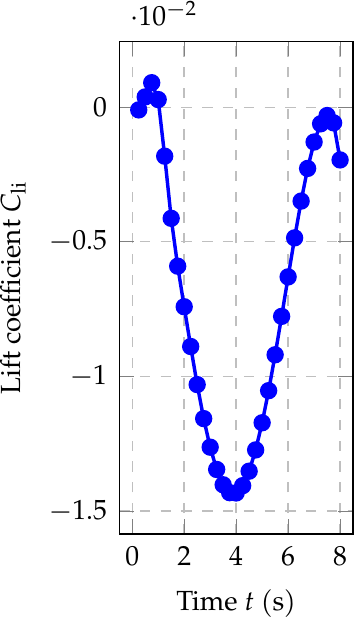}\hfill
\includegraphics{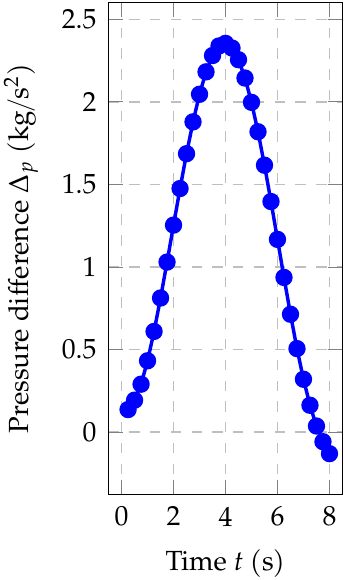}
\includegraphics{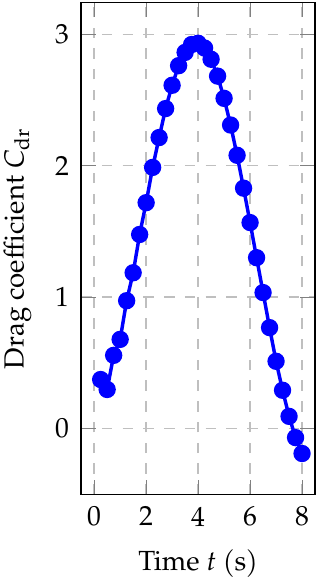}\hfill
\includegraphics{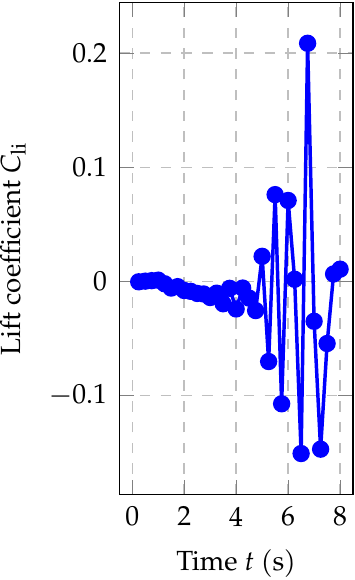}\hfill
\includegraphics{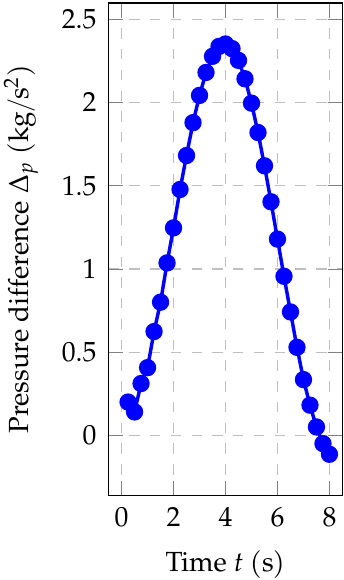}
\caption{Drag and lift coefficient, and pressure difference for $\nu=0.001\ (\text{m}/\text{s})$ for reference (top), $S_1$ (middle) and $S_2$ (bottom).}
\label{f:charnumbers-nu0.001}
\end{figure}

Figure~\ref{f:flowfield} shows the flow field at $t=5.25\ (\text{s})$ for the three different viscosities.
As viscosity decreases, the maximum Reynolds number increases and the flow becomes more turbulent.
While for $\nu=0.1\ (\text{m}/\text{s})$ and $\nu=0.01\ (\text{m}/\text{s})$ the flow is essentially laminar, smaller vortices start to form behind the cylinder for $\nu=0.001\ (\text{m}/\text{s})$.
The resulting evolution over time of the three characteristic numbers for $\nu=0.1\ (\text{m}/\text{s})$ using serial time integration is shown in Figure~\ref{f:charnumbers-nu0.1}.
Both fine integrators $S_1$ and $S_2$ produce essentially identical profiles and their profiles closely match the one generated by the corresponding reference simulation (not shown).
Figure~\ref{f:charnumbers-nu0.01} shows the same profiles for the simulation with $\nu=0.01\ (\text{m}/\text{s})$.
Again, both $S_1$ and $S_2$ produce profiles that match and agree with the results from the corresponding reference (not shown).
Lastly, Figure~\ref{f:charnumbers-nu0.001} shows three profiles for $\nu=0.001\ (\text{m}/\text{s})$, one for the reference simulation, one for $S_1$ and one for $S_2$.
While the drag coefficient and pressure difference agree across all three configurations, $S_1$ produces a lift coefficient profile that is distinctly different from the corresponding reference and $S_2$.
The relatively high numerical diffusion of $S_1$ in combination with a rather low spatial resolution probably prevents $S_1$ from correctly capturing the more turbulent dynamics in this case.
In contrast, although $S_2$ fails to fully reproduce the frequency of oscillations, it still achieves a qualitatively correct representation of the dynamics of the corresponding reference simulation.

\subsection{Convergence of Parareal}
\begin{figure}[!t]
\centering
\includegraphics{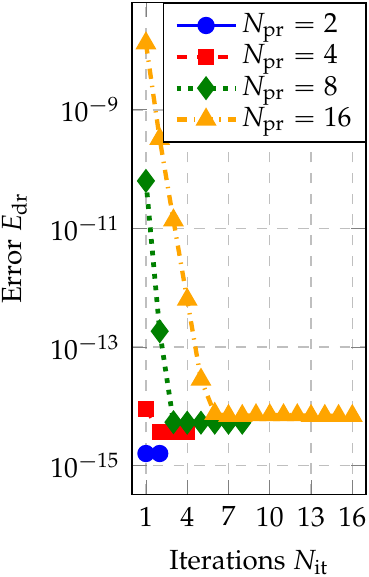}\hfill
\includegraphics{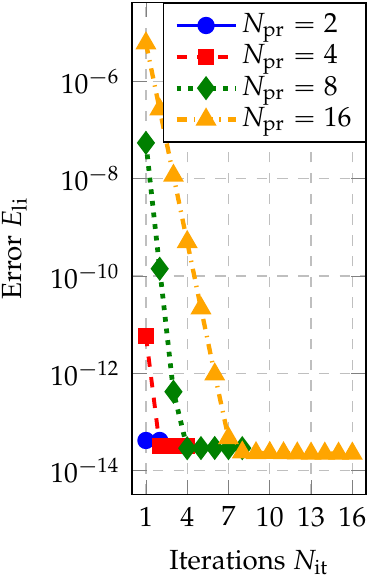}\hfill
\includegraphics{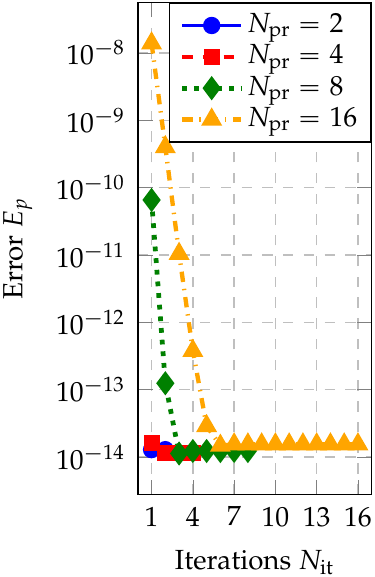}
\includegraphics{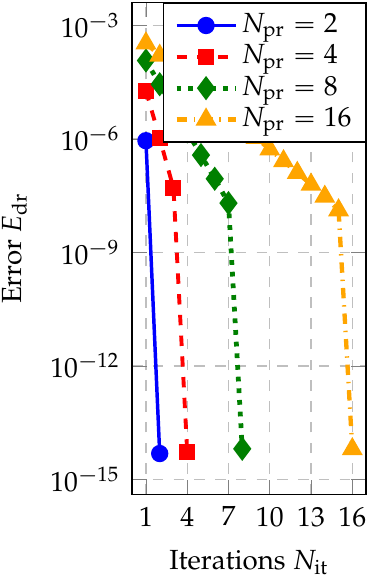}\hfill
\includegraphics{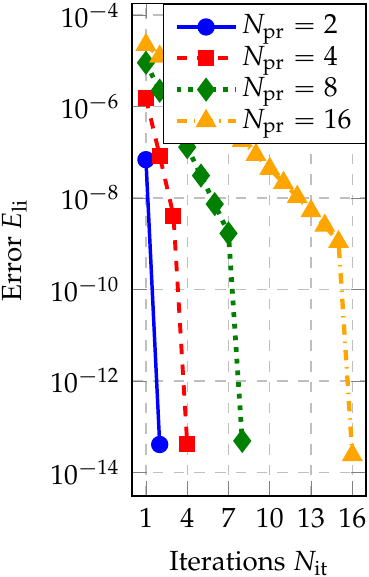}\hfill
\includegraphics{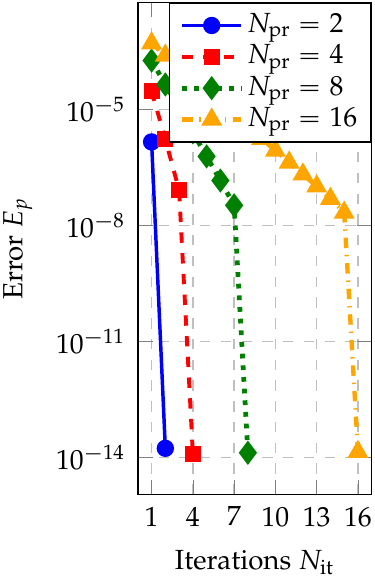}				
\caption{Fine solver convergence errors for drag and lift coefficient, and pressure difference for $\nu=0.1\ (\text{m}/\text{s})$ and $S_1$ (top) and $S_2$ (bottom).}
\label{f:conv-nu0.1}
\end{figure}

Here, we analyse how accurately Parareal reproduces the three characteristic values studied above.
Figure~\ref{f:conv-nu0.1} shows the defect or error according to Equation~\eqref{e:error_norm} in the characteristic values accumulated over all time slices versus the number of iterations.
Here, defect refers to the difference between the solution computed by Parareal and the solution computed by running the fine integrator serially.
For $S_1$, the error for all three quantities, \textit{i.e.} drag coefficient, lift coefficient and pressure difference, quickly goes to zero, that is Parareal rapidly produces values identical to ones obtained from the serial simulation.
As the number of time slices is increased, convergence becomes slower but the increase is not drastic: for $N_{\text{pr}} = 16$ after seven iterations all three characteristic values have converged up to round-off error.
In a production run, where the main goal is to push the defect from Parareal below the discretisation error (see the discussion in~\cite{ArteagaEtAl2015}), significantly fewer iterations will likely suffice.

Decreasing viscosity and thus increasing the Reynolds number range does negatively affect convergence.
For $\nu=0.001\ (\text{m}/\text{s})$, the simulation with $N_{\text{pr}} = 16$ time slices already requires $13$ iterations to converge up to round-off error.
Depending on the desired accuracy, speedup is still possible here but parallel efficiency will likely be lower than in the more laminar case.
Since $S_1$ fails to resolve the full dynamics of the problem, its convergence behaviour is probably not representative of the actual physical dynamics.

This is supported by the fact that for $S_2$ and $\nu=0.001\ (\text{m}/\text{s})$, Parareal essentially no longer converges.
Since $S_2$ does resolve the turbulent dynamics at least partially, in contrast to $S_1$, this suggests that the good convergence of Parareal for $S_1$ and $\nu=0.001\ (\text{m}/\text{s})$ is an artefact produced by excessive numerical diffusion.
The dynamics of the numerical solution are more laminar than they should be, leading to an unrealistic convergence behaviour of Parareeal.
Supposedly, when using $S_1$ on a significantly finer spatial and temporal mesh, a similar deterioration of convergence would be observed, as the numerical solution better resolves the turbulent features of the flow.

Interestingly, this difference between $S_1$ and $S_2$ can already be seen for $\nu=0.1\ (\text{m}/\text{s})$, where the physical dynamics are still quite laminar as well.
Although Parareal for $S_2$ does converge, particularly for larger numbers of time slices, its rate of convergence is lower than for $S_1$.
The benefit of using an integrator with damping properties as coarse integrator has been pointed out before by~\cite{Bal2005} but apparently numerical diffusion from the fine method does help Parareal convergence, too.
Since here the fine method realistically represents the flow features, using a diffusive integrator as fine method in Parareal when simulating turbulent flow could be an easy way to obtain decent convergence.

\begin{figure}[!hpt]
\centering
\includegraphics{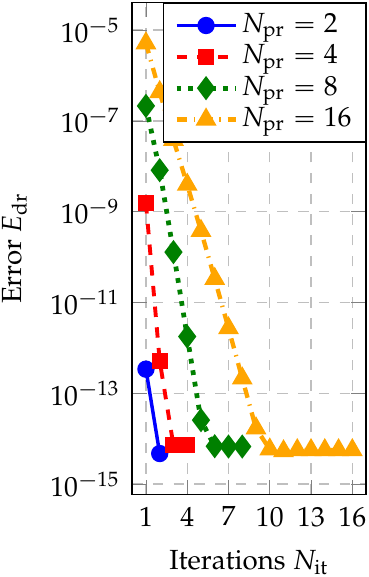}\hfill
\includegraphics{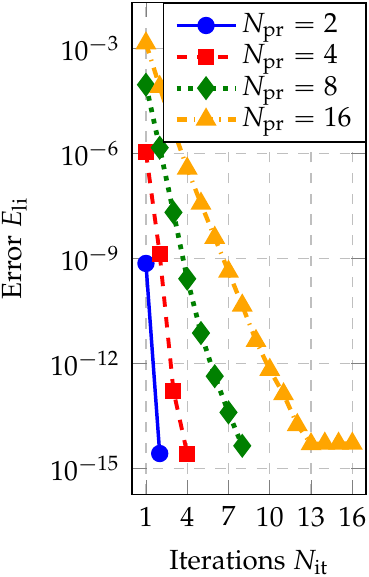}\hfill
\includegraphics{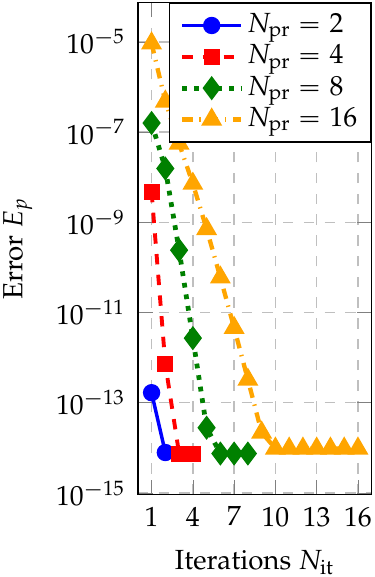}
\includegraphics{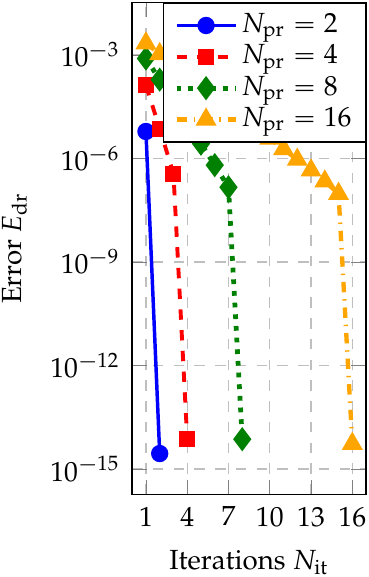}\hfill
\includegraphics{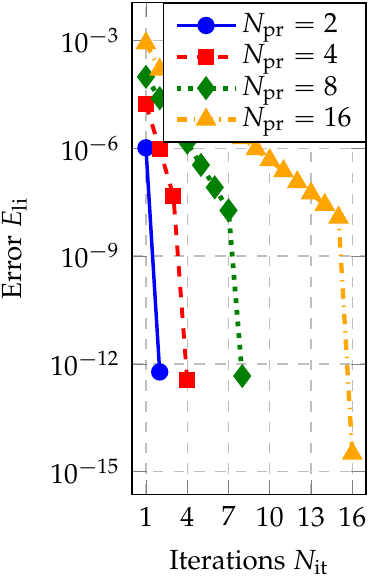}\hfill
\includegraphics{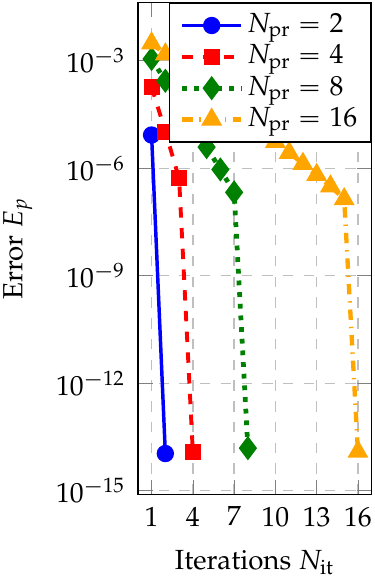}
\caption{Fine solver convergence errors for drag and lift coefficient, and pressure difference for $\nu=0.01\ (\text{m}/\text{s})$ for $S_1$ (top) and $S_2$ (bottom).}
\label{f:conv-nu0.01}
\end{figure}

\begin{figure}[!th]
\centering
\includegraphics{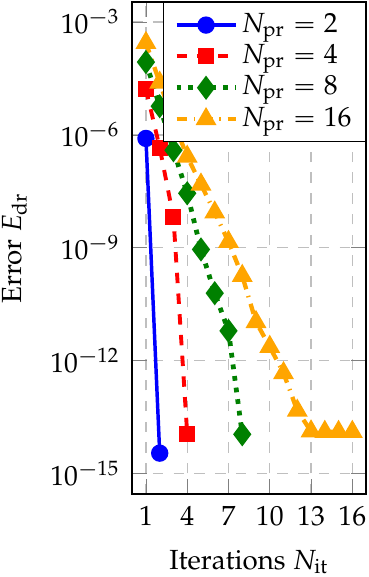}\hfill
\includegraphics{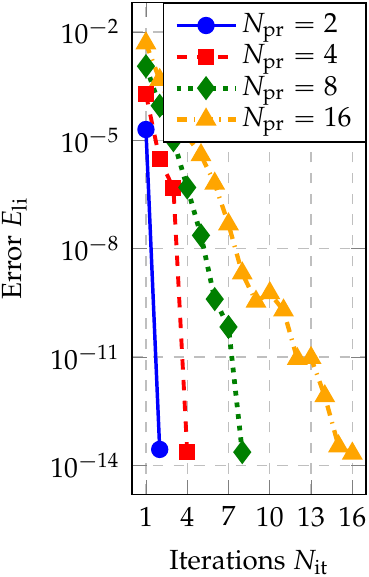}\hfill
\includegraphics{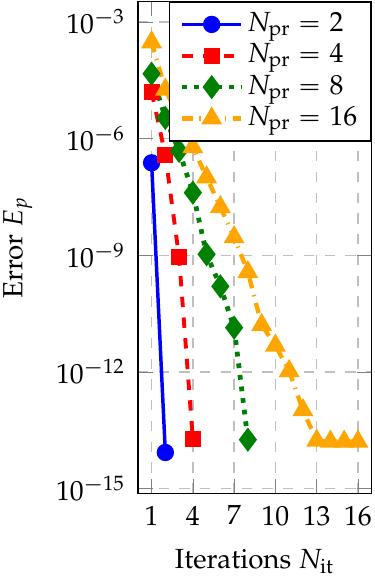}
\includegraphics{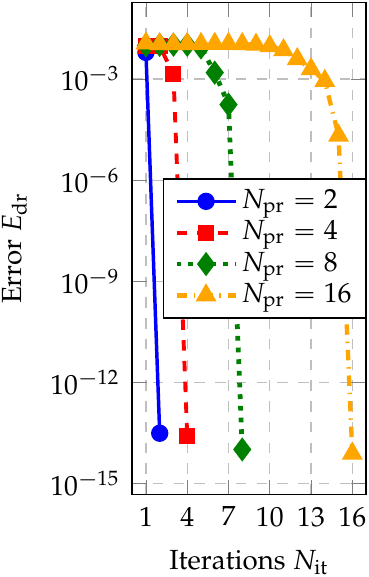}\hfill
\includegraphics{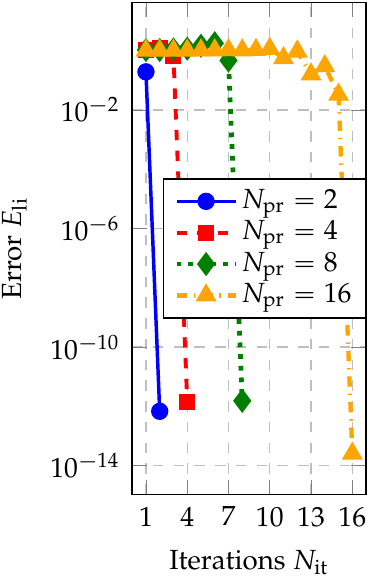}\hfill
\includegraphics{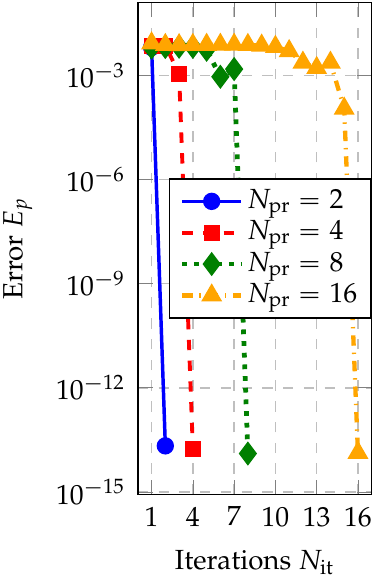}
\caption{Fine solver convergence errors for drag and lift coefficient, and pressure difference for $\nu=0.001\ (\text{m}/\text{s})$ for $S_1$ (top) and $S_2$ (bottom).}
\label{f:conv-nu0.001}
\end{figure}

\section{Conclusions}\label{s:conclusion}
This report extends the investigation by~\cite{SteinerEtAl2015} about how a decreasing viscosity in the Navier-Stokes equations affects the convergence behaviour of the Parareal parallel-in-time method.
An unsteady 2D flow around a cylinder is used as a benchmark.
Two different configurations are tested, one using an implicit Euler as coarse and fine method, the other an implicit Euler as coarse but a second-order fractional step integrator as fine method.
For larger viscosities, both base methods correctly reproduce the evolution of characteristic quantities like drag and lift coefficient and pressure difference.
However, the numerical diffusion from the backward Euler method as fine integrator leads to significantly better convergence of Parareal.

As viscosity decreases, convergence of Parareal becomes slower similarly to the results in~\cite{SteinerEtAl2015}.
However, while for the fractional step method Parareal stalls completely, the backward Euler retains reasonable convergence even for small viscosities.
A comparison of the profiles for characteristic numbers with a reference solution suggests that the good convergence for the backward Euler is likely artificial.
The fine integrator alone captures the relatively smooth profiles of the drag coefficient and pressure difference quite well.
However, the high frequency oscillations in the lift coefficient, which are clearly seen in serial runs using the fractional step integrator, are not present when using backward Euler.
Most likely, the rather high numerical diffusion leads to an artificially laminar flow, so that the good convergence of Parareal is not representative of the used viscosity parameter.
Clearly, when assessing Parareal's convergence for flow problems, care must be taken to ensure that the fine base method correctly resolves the important features of the flow.
Intrinsic convergence of Parareal alone is not a reliable indicator.

There are several works proposing strategies to stabilise Parareal for advection-dominated problems, \textit{e.g.} by~\cite{FarhatCortial2006},~\cite{GanderPetcu2008},~\cite{RuprechtKrause2012} or~\cite{ChenEtAl2014}.
An interesting continuation of the work presented here would be to analyse whether these strategies improve convergence of Parareal for turbulent flows.

\section*{Acknowledgments}

We would like to thank Ernesto Casartelli and Luca Mangani from the Lucerne University of Applied Sciences and Arts (HSLU) in Switzerland for discussions.

The research of A.K., D.R., and R.K. is funded through the ``FUtuRe SwIss Electrical InfraStructure'' (FURIES) project of the Swiss Competence Centers for Energy Research (SCCER) at the Commission for Technology and Innovation (CTI) in Switzerland. The research is also funded by the Deutsche Forschungsgemeinschaft (DFG) as part of the ``ExaSolvers'' project in the Priority Programme 1648 ``Software for Exascale Computing'' (SPPEXA) and by the Swiss National Science Foundation (SNSF) under the lead agency agreement as grant SNSF-145271. 

%
\bibliographystyle{plainnat}
\bibliography{pint,biblio} 

\begin{thebibliography}{25}
\providecommand{\natexlab}[1]{#1}
\providecommand{\url}[1]{\texttt{#1}}
\expandafter\ifx\csname urlstyle\endcsname\relax
  \providecommand{\doi}[1]{doi: #1}\else
  \providecommand{\doi}{doi: \begingroup \urlstyle{rm}\Url}\fi

\bibitem[Arteaga et~al.(2015)Arteaga, Ruprecht, and Krause]{ArteagaEtAl2015}
A.~Arteaga, Daniel Ruprecht, and Rolf Krause.
\newblock {A stencil-based implementation of Parareal in the {C++} domain
  specific embedded language {STELLA}}.
\newblock \emph{Applied Mathematics and Computation}, 2015.
\newblock URL \url{http://dx.doi.org/10.1016/j.amc.2014.12.055}.

\bibitem[Bal(2005)]{Bal2005}
Guillaume Bal.
\newblock {On the convergence and the stability of the parareal algorithm to
  solve partial differential equations}.
\newblock In Ralf Kornhuber and {et al.}, editors, \emph{{Domain Decomposition
  Methods in Science and Engineering}}, volume~40 of \emph{{Lecture Notes in
  Computational Science and Engineering}}, pages 426--432, Berlin, 2005.
  Springer.
\newblock URL \url{http://dx.doi.org/10.1007/3-540-26825-1_43}.

\bibitem[Celledoni and Kvamsdal(2009)]{Celledoni2009}
E.~Celledoni and T.~Kvamsdal.
\newblock {Parallelization in time for thermo-viscoplastic problems in
  extrusion of aluminium}.
\newblock \emph{International Journal for Numerical Methods in Engineering},
  79\penalty0 (5):\penalty0 576--598, 2009.
\newblock URL \url{http://dx.doi.org/10.1002/nme.2585}.

\bibitem[Chen et~al.(2014)Chen, Hesthaven, and Zhu]{ChenEtAl2014}
Feng Chen, Jan~S. Hesthaven, and Xueyu Zhu.
\newblock {On the Use of Reduced Basis Methods to Accelerate and Stabilize the
  Parareal Method}.
\newblock In Alfio Quarteroni and Gianluigi Rozza, editors, \emph{{Reduced
  Order Methods for Modeling and Computational Reduction}}, volume~9 of
  \emph{{MS\&A - Modeling, Simulation and Applications}}, pages 187--214.
  Springer International Publishing, 2014.
\newblock URL \url{http://dx.doi.org/10.1007/978-3-319-02090-7_7}.

\bibitem[Croce et~al.(2014)Croce, Ruprecht, and Krause]{CroceEtAl2014}
Roberto Croce, Daniel Ruprecht, and Rolf Krause.
\newblock {Parallel-in-Space-and-Time Simulation of the Three-Dimensional,
  Unsteady {N}avier-{S}tokes Equations for Incompressible Flow}.
\newblock In Hans~Georg Bock, Xuan~Phu Hoang, Rolf Rannacher, and Johannes~P.
  Schlöder, editors, \emph{{Modeling, Simulation and Optimization of Complex
  Processes -- {HPSC} 2012}}, pages 13--23. Springer International Publishing,
  2014.
\newblock URL \url{http://dx.doi.org/10.1007/978-3-319-09063-4_2}.

\bibitem[{Dongarra et al.}(2014)]{DongarraEtAl2014}
Jack {Dongarra et al.}
\newblock {Applied Mathematics Research for Exascale Computing}.
\newblock Technical Report LLNL-TR-651000, Lawrence Livermore National
  Laboratory, 2014.
\newblock URL
  \url{http://science.energy.gov/~/media/ascr/pdf/research/am/docs/EMWGreport.pdf}.

\bibitem[Farhat et~al.(2006)Farhat, Cortial, Dastillung, and
  Bavestrello]{FarhatCortial2006}
Charbel Farhat, Julien Cortial, C.~Dastillung, and H.~Bavestrello.
\newblock {Time-parallel implicit integrators for the near-real-time prediction
  of linear structural dynamic responses}.
\newblock \emph{International Journal for Numerical Methods in Engineering},
  67:\penalty0 697--724, 2006.
\newblock URL \url{http://dx.doi.org/10.1002/nme.1653}.

\bibitem[Fischer et~al.(2005)Fischer, Hecht, and Maday]{FischerEtAl2005}
P.~F. Fischer, F.~Hecht, and Yvon Maday.
\newblock {A parareal in time semi-implicit approximation of the
  {N}avier-{S}tokes equations}.
\newblock In Ralf Kornhuber and {et al.}, editors, \emph{{Domain Decomposition
  Methods in Science and Engineering}}, volume~40 of \emph{{Lecture Notes in
  Computational Science and Engineering}}, pages 433--440, Berlin, 2005.
  Springer.
\newblock URL \url{http://dx.doi.org/10.1007/3-540-26825-1_44}.

\bibitem[Gander(2015)]{Gander2015_Review}
Martin~J. Gander.
\newblock {50 years of Time Parallel Time Integration}.
\newblock In \emph{Multiple Shooting and Time Domain Decomposition}. Springer,
  2015.
\newblock URL
  \url{{http://www.unige.ch/%7Egander/Preprints/50YearsTimeParallel.pdf}}.

\bibitem[Gander and Neumueller(2014)]{Neumueller2014}
Martin~J. Gander and M.~Neumueller.
\newblock {Analysis of a Time Multigrid Algorithm for {DG}-Discretizations in
  Time}.
\newblock 2014.
\newblock URL \url{http://arxiv.org/abs/1409.5254}.

\bibitem[Gander and Petcu(2008)]{GanderPetcu2008}
Martin~J. Gander and M.~Petcu.
\newblock {Analysis of a {K}rylov Subspace Enhanced Parareal Algorithm for
  Linear Problem}.
\newblock \emph{ESAIM: Proc.}, 25:\penalty0 114--129, 2008.
\newblock URL \url{http://dx.doi.org/10.1051/proc:082508}.

\bibitem[Gander and Vandewalle(2007)]{GanderVandewalle2007}
Martin~J. Gander and Stefan Vandewalle.
\newblock {On the Superlinear and Linear Convergence of the Parareal
  Algorithm}.
\newblock In Olof~B. Widlund and David~E. Keyes, editors, \emph{{Domain
  Decomposition Methods in Science and Engineering}}, volume~55 of
  \emph{{Lecture Notes in Computational Science and Engineering}}, pages
  291--298. Springer Berlin Heidelberg, 2007.
\newblock URL \url{http://dx.doi.org/10.1007/978-3-540-34469-8_34}.

\bibitem[John(2004)]{John2004a}
Volker John.
\newblock {Reference values for drag and lift of a two-dimensional
  time-dependent flow around a cylinder}.
\newblock \emph{International Journal for Numerical Methods in Fluids},
  44\penalty0 (7):\penalty0 777--788, Mar 2004.
\newblock \doi{10.1002/fld.679}.

\bibitem[Kreienbuehl et~al.(2015)Kreienbuehl, Naegel, Ruprecht, Speck, Wittum,
  and Krause]{KreienbuehlEtAl2015a}
Andreas Kreienbuehl, Arne Naegel, Daniel Ruprecht, Robert Speck, Gabriel
  Wittum, and Rolf Krause.
\newblock {Numerical simulation of skin transport using Parareal}.
\newblock \emph{Computing and Visualization in Science}, Aug 2015.
\newblock \doi{10.1007/s00791-015-0246-y}.
\newblock URL \url{http://arxiv.org/abs/1502.03645}.

\bibitem[Lions et~al.(2001)Lions, Maday, and Turinici]{LionsEtAl2001}
J.-L. Lions, Yvon Maday, and Gabriel Turinici.
\newblock {A "parareal" in time discretization of {PDE}'s}.
\newblock \emph{Comptes Rendus de l'Académie des Sciences - Series I -
  Mathematics}, 332:\penalty0 661--668, 2001.
\newblock URL \url{http://dx.doi.org/10.1016/S0764-4442(00)01793-6}.

\bibitem[Minion(2010)]{Minion2010}
Michael~L. Minion.
\newblock {A Hybrid Parareal Spectral Deferred Corrections Method}.
\newblock \emph{Communications in Applied Mathematics and Computational
  Science}, 5\penalty0 (2):\penalty0 265--301, 2010.
\newblock URL \url{http://dx.doi.org/10.2140/camcos.2010.5.265}.

\bibitem[Ruprecht and Krause(2012)]{RuprechtKrause2012}
Daniel Ruprecht and Rolf Krause.
\newblock {Explicit parallel-in-time integration of a linear acoustic-advection
  system}.
\newblock \emph{Computers \& Fluids}, 59\penalty0 (0):\penalty0 72--83, 2012.
\newblock URL \url{http://dx.doi.org/10.1016/j.compfluid.2012.02.015}.

\bibitem[Ruprecht et~al.(2013)Ruprecht, Speck, Emmett, Bolten, and
  Krause]{RuprechtEtAl2013_SC}
Daniel Ruprecht, Robert Speck, Matthew Emmett, Matthias Bolten, and Rolf
  Krause.
\newblock {Poster: Extreme-scale space-time parallelism}.
\newblock In \emph{{Proceedings of the 2013 Conference on High Performance
  Computing Networking, Storage and Analysis Companion}}, {SC '13 Companion},
  2013.
\newblock URL
  \url{http://sc13.supercomputing.org/sites/default/files/PostersArchive/tech_posters/post148s2-file3.pdf}.

\bibitem[Sch{\"a}fer et~al.(1996)Sch{\"a}fer, Turek, Durst, Krause, and
  Rannacher]{SchaeferAndTurek1996a}
Michael Sch{\"a}fer, Stefan Turek, Franz Durst, Egon Krause, and Rolf
  Rannacher.
\newblock {Benchmark Computations of Laminar Flow Around a Cylinder}.
\newblock In Ernst~Heinrich Hirschel, editor, \emph{{Flow Simulation with
  High-Performance Computers II}}, volume~48 of \emph{Notes on Numerical Fluid
  Mechanics (NNFM)}, pages 547--566. Vieweg\texttt{+}Teubner Verlag, 1996.
\newblock ISBN (13) 9783322898517.
\newblock \doi{10.1007/978-3-322-89849-4\_39}.

\bibitem[Speck et~al.(2012)Speck, Ruprecht, Krause, Emmett, Minion, Winkel, and
  Gibbon]{SpeckEtAl2012}
Robert Speck, Daniel Ruprecht, Rolf Krause, Matthew Emmett, Michael~L. Minion,
  Mathias Winkel, and Paul Gibbon.
\newblock {A massively space-time parallel {N}-body solver}.
\newblock In \emph{{Proceedings of the International Conference on High
  Performance Computing, Networking, Storage and Analysis}}, {SC '12}, pages
  92:1--92:11, Los Alamitos, CA, USA, 2012. IEEE Computer Society Press.
\newblock URL \url{http://dx.doi.org/10.1109/SC.2012.6}.

\bibitem[Steiner et~al.(2015)Steiner, Ruprecht, Speck, and
  Krause]{SteinerEtAl2015}
J.~Steiner, Daniel Ruprecht, Robert Speck, and Rolf Krause.
\newblock {Convergence of {P}arareal for the {N}avier-{S}tokes equations
  depending on the {R}eynolds number}.
\newblock In Assyr Abdulle, Simone Deparis, Daniel Kressner, Fabio Nobile, and
  Marco Picasso, editors, \emph{{Numerical Mathematics and Advanced
  Applications - ENUMATH 2013}}, volume 103 of \emph{{Lecture Notes in
  Computational Science and Engineering}}, pages 195--202. Springer
  International Publishing, 2015.
\newblock URL \url{http://dx.doi.org/10.1007/978-3-319-10705-9_19}.

\bibitem[Trindade and Pereira(2004)]{Trindade2004}
J.~M.~F. Trindade and J.~C.~F. Pereira.
\newblock {Parallel-in-time simulation of the unsteady {N}avier-{S}tokes
  equations for incompressible flow}.
\newblock \emph{International Journal for Numerical Methods in Fluids},
  45\penalty0 (10):\penalty0 1123--1136, 2004.
\newblock URL \url{http://dx.doi.org/10.1002/fld.732}.

\bibitem[Trindade and Pereira(2006)]{Trindade2006}
J.~M.~F. Trindade and J.~C.~F. Pereira.
\newblock {Parallel-in-Time Simulation of Two-Dimensional, Unsteady,
  Incompressible Laminar Flows}.
\newblock \emph{Numerical Heat Transfer, Part B: Fundamentals}, 50\penalty0
  (1):\penalty0 25--40, 2006.
\newblock URL \url{http://dx.doi.org/10.1080/10407790500459379}.

\bibitem[Vogel(2014)]{VogelDiss2014}
Andreas Vogel.
\newblock \emph{{Flexible und kombinierbare Implementierung von
  Finite-Volumen-Verfahren h\"oherer Ordnung mit Anwendungen für die
  Konvektions-Diffusions-, Navier-Stokes- und Nernst-Planck-Gleichungen sowie
  dichtegetriebene Grundwasserströmung in por\"osen Medien}}.
\newblock PhD thesis, Johann Wolfgang Goethe-Universit\"at Frankfurt, 2014.

\bibitem[Vogel et~al.(2013)Vogel, Reiter, Rupp, N\"agel, and
  Wittum]{VogelUG42013}
Andreas Vogel, Sebastian Reiter, Martin Rupp, Arne N\"agel, and Gabriel Wittum.
\newblock {UG 4: A novel flexible software system for simulating PDE based
  models on high performance computers}.
\newblock \emph{Computing and Visualization in Science}, 16\penalty0
  (4):\penalty0 165--179, Aug 2013.
\newblock ISSN 1433-0369.
\newblock \doi{10.1007/s00791-014-0232-9}.
\newblock URL \url{http://dx.doi.org/10.1007/s00791-014-0232-9}.

\end{thebibliography}



\end{document}